\documentclass[twocolumn,english]{article}
\usepackage[T1]{fontenc}
\usepackage[latin1]{inputenc}
\usepackage{graphicx}

\makeatletter

\providecommand{\LyX}{L\kern-.1667em\lower.25em\hbox{Y}\kern-.125emX\@}
 \newcommand{\lyxrightaddress}[1]{
   \par {\raggedleft \begin{tabular}{l}\ignorespaces
   #1
   \end{tabular}
   \vspace{1.4em}
   \par}
 }

\usepackage{babel}
\makeatother
\begin{document}
%\begin{minipage}[c]{1.0\textwidth}%
\twocolumn[
%\begin{@twocolumnfalse}
\lyxrightaddress{SLAC-PUB-10154\\
$BaBar$-PROC-02/096\\
hep-ex/xxxxxxx\\
July, 2002}

\textbf{\large $CP$ Violation, Mixing and Lifetime Results from}
\textbf{\emph{\large BaBar}}{\large \par}	
\medskip{}	
D. Payne\footnotemark (For the $BaBar$ collaboration)
\medskip{}{\par}
Department of Physics,Oliver Lodge, University of Liverpool,L69 7ZE,
Liverpool, U.K.

\bigskip{}
{\small The $BaBar$ collaboration has analysed 60M $B\overline{B}$
pairs collected at the $\Upsilon (4S)$ resonance at the PEP II asymetric
collider at SLAC. Using this data sample we have measured the $CP$
violation parameters $\sin (2\beta )=0.75\pm 0.09_{STAT}\pm 0.04_{SYST}$
and $|\lambda |=0.92\pm 0.06_{STAT}\pm 0.02_{SYST}$ from $B^{0}\rightarrow c\overline{c}+K^{0(*)}+c.c.$
decays. From charmless 2-body $B$ decays we measure $A_{K\pi }=-0.05\pm 0.06\pm 0.01(-0.14,+0.05)$,$S_{\pi \pi }=-0.01\pm 0.37\pm 0.07(-0.66,+0.62)$,
$C_{\pi \pi }=-0.02\pm 0.29\pm 0.07(-0.54,+0.48)$. A number of $B$ lifetime and mixing parameters, extracted from subsamples of this data set, are also presented.}{\small \par}

\bigskip{}
Presented at QCD 02 Montpellier, France 2-9 July 2002.
\bigskip{}
\vspace*{0.25in}
%\end{minipage}%
]
%\end{@twocolumnfalse}

\footnotetext{Work supported in part by the Department of Energy contract DE-AC03-76SF00515}

\bigskip{}
\section{Introduction}

\emph{B} factories like PEP II produce vast quantities of \emph{B}
pairs, and allow the properties of the \emph{B} meson to be studied
in unprecedented detail. Of particular interest are the $CP$ violating
aspects of its behavior. $CP$ violation in the $b$ sector was first
observed at BaBar in the summer of 2001 with the measurement of $\sin (2\beta )$
from $B^{0}\rightarrow c\overline{c}+K^{0(*)}+c.c.$ \cite{babar-s2b-1}.
With increasing datasets, $\sin (2\beta )$ can be measured with ever
greater accuracy. Also, we can move on to our next goal in the study
of $CP$ violation in the $b$ sector, the measurement of the parameter
$\alpha $.

In addition, many other aspects of $B$ meson behavior can be studied,
such as $B$ lifetime and $B^{0}\overline{B^{0}}$ mixing. The methods
needed to extract values for these quantities are similar to those
used in $CP$ analysis, and many techniques are common to both.

\section{Time Evolution of the $B$}

\label{sec:Time-Evolution}For charged $B$s, evolution with time
is a simple matter of exponential decay (decay amplitude $f(\Delta t)=\frac{1}{\tau _{B^{\pm }}}e^{-\Delta t/\tau _{B^{\pm }}}$).
Neutral $B$s can mix, changing flavor $B^{0}\rightarrow \overline{B^{0}}+c.c.$,
complicating their behavior ( $f_{\pm }(\Delta t)=\frac{1}{4\tau _{B^{0}}}e^{-\Delta t/\tau _{B^{0}}}\times (1\pm \cos (\Delta m_{d}\Delta t))$,
where +/- stands for $B^{0}/\overline{B^{0}}$ and $\Delta m_{d}$
is the mass difference between the heavy and light mass eigenstates).
Where $CP$ violation occurs (in decays to appropriate $CP$ eigenstates)
the decay amplitude is described by $f(\Delta t)_{\pm }=\frac{1}{4\tau _{B^{0}}}e^{-\Delta t/\tau _{B^{0}}}\times (1\pm S\sin (\Delta m_{d}\Delta t)\pm C\cos (\Delta m_{d}\Delta t))$,
where $S$ and $C$ are constants which depend upon the final $CP$
eigenstates. $C$ is a measure of the direct $CP$ violation ($CP$
violation in decay) occuring in the given decay, and $S$ is determined
by $CP$ violation from the interference between mixing and decay.

\section{The Dataset}

The data used were collected with the BaBar detector. Different datasets
were used with different analysis: the $CP$ results use approximately
$56fb^{-1}$ of data taken at the $\Upsilon (4S)$ resonance, the
Mixing result using fully reconstructed hadronic events shown here
uses $30fb^{-1}$ and other results use $20fb^{-1}$.

The BaBar detector is described in detail elsewhere \cite{NIM}.

\section{Common Techniques}

There are several general methods used in some or all of the analysis
that will be discussed here.

\subsection{Measurement of $\Delta t$}

At BaBar, the measurement of $\Delta t$ (the decay time difference
between the first and second B to decay) relies heavily on the boost
to the centre of mass brought about by PEP II's asymmetric design.
$B$s produced in pairs from the decay of the $\Upsilon (4S)$ resonance
are almost at rest in the centre of mass. Thanks to the boost, they
fly almost straight down the z axis of the detector. If the boost
and the separation in $z$ of the 2 $B$s decay vertices are known,
then $\Delta t$ is known (to within small corrections).

Where a $B$ is reconstructed, a least $\chi ^{2}$ fit is formed
on all the tracks used in that reconstruction to find the vertex of
that $B$. To find the vertex of the other $B$ in the event, all
remaining tracks (those not used in the reconstruction) also have
a least $\chi ^{2}$ fit performed on them. Since decays of the $\Upsilon (4S)$
produce two $B$s and nothing else, they must form the vertex of the
other $B$. Tracks which contribute too large a value to the $\chi ^{2}$
are removed, as are those thought likely to come from $K_{s},\Lambda $
or conversions. 

The average separation of the $B$ vertices is $250\mu m$. Where
a $B$ is reconstructed, is vertex position can be measured with an
error of $70\mu m$, where a $B$ vertex is formed from leftover tracks
the error is $180\mu m$. The error on $\Delta t$ is therefore a
significant fraction of $\Delta t$. As a result in fits the $\Delta t$
dependent decay rate must be convoluted with a resolution function.
At $BaBar$ we use resolution functions paramaterised in terms of
event by event errors. The exact model used varies between analysis.

\subsection{Flavor tagging}

Mixing and $CP$ measurements require knowledge of the flavor of the
decaying $B$s in addition to $\Delta t$. In some cases, where a
$B$ is reconstructed its flavor can be deduced from its mode of decay.
To measure the flavor of an \emph{un}reconstructed $B$, we infer
it from its decay products using leftover tracks in the event. Several
different methods are used. If a high energy lepton is observed amongst
the decay products, the flavor of the $B$ is very accurately determined
by its charge ($l^{-}\Rightarrow B^{0}$ + c.c.). The presence of
a charged Kaon is also a tell tale ($K^{-}\Rightarrow B^{0}$ + c.c.).
If neither of these methods give an answer, we use a neural net technique
(which gathers its information mostly from the charge of slow pions
and unidetified leptons).

These methods are only of finite accuracy, and when flavor is measured
incorrectly it leads to a diluting effect on the final measurement.
Fortunately, this dilution can be measured in the process of performing
a mixing fit, and this measured dilution can then be used as a correction
factor in measurements of $CP$ asymmetries.

\subsection{Samples}

\label{sub:Samples}
\begin{figure}
\includegraphics[  scale=0.4]{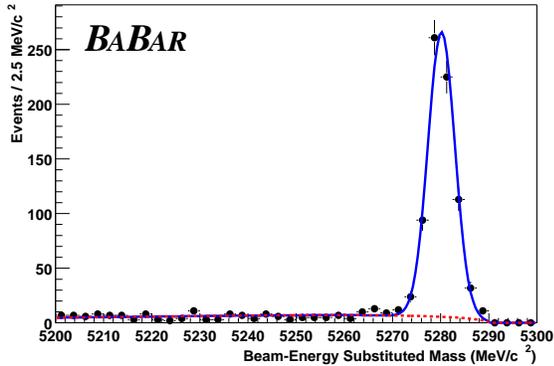}

\caption{\label{cap:Jpsi-Ks-smashing}$M_{es}$distributions for $B^{0}\rightarrow c\overline{c}K_{s}$
. Fit is Gaussian for signal, Argus (phase space) for combinatoric
background.}
\end{figure}

The $B^{0}\rightarrow c\overline{c}+K^{0(*)}+c.c.$
sample is of particular importance for measurement of $\sin (2\beta )$.
In this sample, one of the $B$s can be fully reconstructed and be
seen to have decayed to Charmonium ($J/\psi $, $\psi (2S)$, $\chi _{c1}$)
and to either $K_{s}$ ($CP$ odd, $\eta =-1$) or $K_{L}$($CP$
even, $\eta =1$). A plot of the energy substituted $B$ mass from
this sample is shown in Fig. \ref{cap:Jpsi-Ks-smashing}. The Charmless
two body $B$ decay sample uses the $CP$ eigenstates $B^{0}\rightarrow \pi \pi ,K\pi $
and $KK$.

The fully reconstructed hadronic sample uses decays that identify
the flavor of the decaying $B$. We use $\overline{B^{0}}\rightarrow D^{(*)+}\pi ^{-},D^{(*)+}\rho ^{-},D^{(*)+}a1^{-},J/\psi \overline{K^{*0}}+c.c$
and $B^{-}\rightarrow D^{(*)0}\pi ^{-},J/\psi K^{-},\psi (2S)K^{-}+c.c.$,
where $D^{*+}\rightarrow D^{0}\pi ^{+}$, $D^{*0}\rightarrow D^{0}\pi ^{0}$,
$D^{0}\rightarrow K^{-}\pi ^{+},K^{-}\pi ^{+}\pi ^{0},K_{S}^{0}\pi ^{+}\pi ^{-},K^{-}\pi ^{+}\pi ^{-}\pi ^{+}$,
$D^{+}\rightarrow K^{-}\pi ^{+}\pi ^{+},K_{s}^{0}\pi ^{+}$ , $\overline{K^{*0}}\rightarrow K^{-}\pi ^{+}$
(all + c.c.). These provide enormous $(10^{4})$ event samples.

Greater efficiency can be provided by partial reconstruction techniques,
albeit with correspondingly poorer purity. To form the partially reconstructed
$B^{0}\rightarrow D^{*-}\rho ^{+},D^{*-}\pi ^{+}$ sample, no attempt
is made to reconstruct the $D^{0}$ that the $D^{*}$ decays to, instead
its mass is deduced using the tightly constrained kinematics that
result from working at the $\Upsilon (4S)$. Likewise the partially
reconstructed semi-leptonic $B$ sample chooses events with a fast
lepton and a slow pion that are consistent with $B^{0}\rightarrow D^{*}l\nu $
decay. The dilepton analysis takes this to its extreme by requiring
only two fast leptons (looking for cases where both $B$s have decayed
semi-leptonic+ly, $1\%$ of $B$ decays).

\section{$B^{0}$and $B^{\pm }$ Lifetimes}

A number of $B$ lifetime measurements have been made at $BaBar$.
Some are presented here. From hadronic, fully reconstructed $B$ decays
$\tau _{B^{0}}=1.546\pm 0.032\pm 0.022ps$ , $\tau _{B^{\pm }}=1.673\pm 0.032\pm 0.023ps$
,$\tau _{B^{0}}/\tau _{B^{\pm }}=1.082\pm 0.026\pm 0.012$. Here the
resolution function is the largest systematic uncertainty. From partially
reconstructed $D^{*}\pi $ $\tau _{B^{0}}=1.510\pm 0.040\pm 0.038ps$
and from partially reconstructed $D^{*}\rho $ $\tau _{B^{0}}=1.616\pm 0.064\pm 0.075ps$
. In this case the systematic error is dominated by the lack of Monte
Carlo statistics and the constitution of the background. Partially
reconstructed semi-leptonic $B$ decays give a value of $\tau _{B^{0}}=1.529\pm 0.012\pm 0.029ps$
with the largest systematic errors being the impact of the choice
of resolution model . The $BaBar$ dilepton analysis gives $\tau _{B^{0}}=1.557\pm 0.028\pm 0.027ps$,$\tau _{B^{\pm }}=1.655\pm 0.026\pm 0.027ps$,$\tau _{B^{0}}/\tau _{B^{\pm }}=1.064\pm 0.031\pm 0.026$,
and in this analysis the resolution and background models dominate
the systematic error.

\section{$B^{0}\overline{B^{0}}$ Mixing}

$BaBar$ has made a number of measurements of $B^{0}$ mixing. From
fully reconstructed hadronic $B$ decays, $\Delta m_{d}=0.516\pm 0.016\pm 0.010ps^{-1}$
with the uncertainty in the value of the $B^{0}$ lifetime (taken
from \cite{PDG}) an the possible effects of tracking misalignment
dominating the error. The dilepton analysis gives a value of $\Delta m_{d}=0.493\pm 0.012\pm 0.009ps^{-1}$.
Again, the $B^{0}$ lifetime is a significant source of systematic
error, as is the parameterization of the resolution function.

\section{$\sin (2\beta )$ from $B^{0}\rightarrow c\overline{c}+K^{0(*)}$}

This analysis uses the sample of $B^{0}\rightarrow $Charmonium $K_{s}$
(or $K_{l}$ or $K^{*0}$) + c.c. $CP$ eigenstates described in section
\ref{sub:Samples}. For these $CP$ eigenstates, the decay is predicted
to be dominated by a single tree diagram, making direct $CP$ violation
an insignificant effect. In terms of the formalism described in section
\ref{sec:Time-Evolution}, this means that $C=0$. Also, for these
decays, $S=\sin (2\beta )$, making a measurement of time dependent
$CP$ violation in decays to these eigenstates a theoretically clean
measurement of the $CKM$ parameter $\beta $\cite{Neubert}.

In addition to the sample of decays to $CP$ eigenstates, we use the
hadronic $B$ sample to evaluate the dilution effect from wrongly
measured flavor. To extract the value of $\sin (2\beta )$ whilst
also fitting for the resolution function and the flavor tagging dilution
(both of which may vary depending on the method used to tag the flavor)
we use a 34 parameter fit. We fix $\Delta m_{d}$ and $\tau _{B^{0}}$
to the values quoted in \cite{PDG}.

\begin{figure}
\includegraphics[  bb=0 0 425bp 638bp,
  scale=0.35]{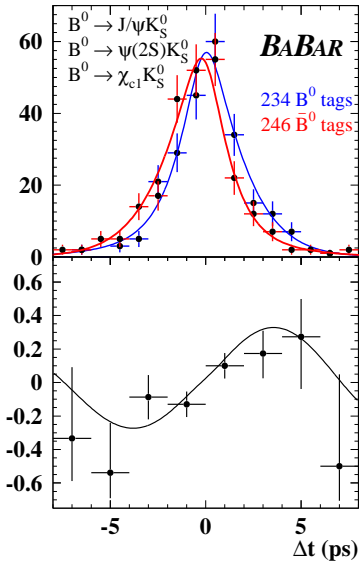}

\caption{\label{cap:Delta-t}$\Delta t$ for $B^{0}$and $\overline{B^{0}}$
(top), $\frac{\Delta t(B^{0})-\Delta t(\overline{B^{0}})}{\Delta t(B^{0})+\Delta t(\overline{B^{0}})}$
(bottom), $\eta =-1$ eigenstates.}
\end{figure}

Using this method, we obtain a value for $\sin (2\beta )$ of $0.75\pm 0.09_{STAT}\pm 0.04_{SYST}$.
This is entirely in line with standard model expectations. Figure
\ref{cap:Delta-t} shows graphicly the asymmetry results. The time
dependent asymmetry can be clearly seen here.

 The chief systematics are the uncertainty in $\Delta m_{d}$ and
$\tau _{B^{0}}$, the properties of the background, and the effect
of possible misalignment of our vertex tracker system. 

To check the assumption that $C=0$, we repeat the fit, allowing for
the addition of a cosine term (allowing $C\neq 0$). We state this
measurement in terms of $|\lambda |$, where $C=(1-|\lambda |^{2})/(1+|\lambda |^{2})$.
We measure $|\lambda |=0.92\pm 0.06_{STAT}\pm 0.02_{SYST}$, consistent
with the absence of direct $CP$ violation in these decays and therefore
in line with the standard model.

\section{$\sin (2\alpha )$ and Direct $CP$ Violation From $B^{0}\rightarrow h^{+}h^{-}$}

It should be possible to extract $\sin (2\alpha )$ from the time
dependant decay rate of $B^{0}\rightarrow \pi ^{+}\pi ^{-}$decays
in much the same way that $\sin (2\beta )$ is measured using $B^{0}\rightarrow c\overline{c}+K^{0(*)}$
decays \cite{Neubert}. $BaBar$ does indeed perform an analysis using
$B^{0}\rightarrow \pi ^{+}\pi ^{-}$ similar to that for used for
$\sin (2\beta )$, but there are some significant differences. In
these decays, we cannot assume that a single diagram dominates. We
would expect both $S_{\pi \pi }$ and $C_{\pi \pi }$ (as defined
in section \ref{sec:Time-Evolution}) to be nonzero, also $\sin (2\alpha )$
is \emph{not} simply $S_{\pi \pi }$, and cannot be trivially extracted.
Instead, we quote results in terms of $S_{\pi \pi }$ and $C_{\pi \pi }$. 

In addition there are technical differences. Background is much higher,
and distinguishing $\pi \pi $ and $K\pi $ is important - misidentification
would distort the result.

The analysis is done in two stages. First, the branching fractions
for $\pi \pi $, $K\pi $ and $KK$ are determined, and $A_{K\pi }$
(direct $CP$ in the $K\pi $ mode) measured. This leads to the results
shown in table \ref{cap:twobodyBF} %
\begin{table}
\begin{tabular}{|c|c|c|}
\hline 
Mode&
Yield (events)&
BF $(10^{-6})$\\
\hline
\hline 
$B^{0}\rightarrow \pi \pi $&
$124_{-15-9}^{+16+7}$&
$5.4\pm 0.7\pm 0.4$\\
\hline 
$B^{0}\rightarrow K\pi $&
$403\pm 24\pm 15$&
$17.8\pm 1.1\pm 0.8$\\
\hline 
$B^{0}\rightarrow KK$&
$<15.6(90\%C.L.)$&
$<1.1(90\%C.L.)$\\
\hline
\end{tabular}

\caption{\label{cap:twobodyBF}Two Body Branching Fractions}
\end{table}
 and the measurement of $A_{K\pi }$ as $-0.05\pm 0.06_{STAT}\pm 0.01_{SYST}$.
Next, a fit is performed for $S_{\pi \pi }$ and $C_{\pi \pi }$,
with the results $S_{\pi \pi }=-0.01\pm 0.37_{STAT}\pm 0.07_{SYST}(-0.66,+0.62)$,
$C_{\pi \pi }=-0.02\pm 0.29_{STAT}\pm 0.07_{SYST}(-0.54,+0.48)$.
No significant $CP$ violation is seen here. The dominant systematic
of this analysis comes from $K/\pi $ separation.

\section{Conclusion}

We have presented a series of measurement of $B$ lifetime, mixing
amplitude and the $CP$ violation parameters $\sin (2\beta )$, $|\lambda |$,
$S_{\pi \pi }$, $C_{\pi \pi }$ and $A_{K\pi }$. $\sin (2\beta )$
is measured with increased accuracy. No $CP$ violation is yet seen
in $B^{0}\rightarrow \pi \pi $ or $K\pi $.

\end{document}